\documentclass[manuscript]{aastex}

\shorttitle{Fermi Bubble} \shortauthors{Cheng et al.}

\begin{document}


\title{Origin of the Fermi Bubble}


\author{ K.-S. Cheng\altaffilmark{1}, D. O. Chernyshov\altaffilmark{1,2,3},
V. A. Dogiel\altaffilmark{1,2}, C.-M. Ko\altaffilmark{3}, W.-H.
Ip\altaffilmark{3}}

\affil{ $^1$ Department of Physics, University of Hong Kong,
Pokfulam Road, Hong Kong, China} \affil{$^2$ I.E.Tamm Theoretical
Physics Division of P.N.Lebedev Institute of Physics, Leninskii
pr. 53, 119991 Moscow, Russia } \affil{$^3$ Institute of
Astronomy, National Central University, Chung-Li 32054, Taiwan }


\begin{abstract}
Fermi has discovered two giant gamma-ray-emitting bubbles that
extend nearly 10kpc in diameter north and south of the galactic
center (GC). The existence of the bubbles was first evidenced in X-rays
detected by ROSAT and later WMAP detected an excess of radio
signals at the location of the gamma-ray bubbles. We propose that
periodic star capture processes by the galactic supermassive black
hole, Sgr A$^*$, with a capture rate $3\times 10^{-5}$yr$^{-1}$
and energy release $\sim 3\times 10^{52}$erg per capture can
produce very hot plasma $\sim 10$keV with a wind velocity $\sim
10^8$cm/s injected into the halo and heat up the halo gas to $\sim
1$keV, which produces thermal X-rays. The periodic injection of
hot plasma can produce shocks in the halo and accelerate electrons
to $\sim$TeV, which produce radio emission via
synchrotron radiation, and gamma-rays via inverse Compton
scattering with the relic and the galactic soft photons.

\end{abstract}



%
%


\section{Introduction}

Observations reveal many evidences of unusual processes occurring
in the region of GC. For instance, the enigmatic 511 keV
annihilation emission discovered by INTEGRAL \citep[see
e.g.][]{knoed} whose origin is still debated, the hot plasma with
 temperature about 10 keV which cannot be confined in the GC
and, therefore, sources with a power about $10^{41}$ erg s$^{-1}$
are required to heat the plasma \citep{koyama07}. In fact plasma outflows with the velocity $\ga
100$ km s$^{-1}$ are observed in the nucleus regions of our Galaxy
\citep{crocker2} and  Andromeda
\citep{gilf}. Time variations of the 6.4 keV line and X-ray
continuum emission observed in the direction of molecular clouds
in the GC which are supposed to be a reflection of a giant X-ray
flare occurred several hundred years ago in the GC \citep{inui,ponti,terr}. HESS observations of the GC
in the TeV energy range indicated an explosive injection of CR
there which might be associated with the supermassive black hole
Sgr A$^\ast$(e.g. Aharonian et al. 2006).

 Recent analysis of
Fermi LAT data \citep[see][]{meng,dobler2} discovered a new
evidence of the GC activity. They found two giant features of
gamma-ray emission in the range 1 -100 GeV, extending 50 degrees
($\sim 10$ kpc) above and below the Galactic center - the Fermi
Bubble (FB). They presented a list of mechanisms that may
contribute
 to the energy release and particle production necessary to
explain the gamma-ray emission from the bubble. They noticed,
however, that most likely the Fermi bubble structure was created
by some large episode of energy injection in the GC, such as a
past accretion event onto the central supermassive black hole
(SMBH) in the last $\sim 10$ Myr. They cast doubt on the idea that
the Fermi bubble was generated by previous starburst activity in
the GC because there was no evidence of massive supernova
explosions ($\sim 10^{4}-10^{5}$) in the past $\sim 10^7$ yr
towards the GC.
Besides, these supernova remnants should be traced by the line
emission of radioactive $^{26}$Al. Observations do not show
significant concentration of $^{26}$Al line towards the GC
\citep{diehl}.

\citet{crocker} and \citet{crocker2,crocker1} argued that the
procedure used by \citet{meng} did not remove contributions of CR
interaction with an ionised gas, then gamma-rays could be produced
by protons interaction with the fully ionised plasma in the halo.
\citet{crocker} also argued that the lifetime of these protons can
be very long because the plasma is extremely turbulent in this
region therefore protons could be trapped there for a time scale
$\tau_{pp}\ga 10^{10}$ yr, then the observed gamma-rays can be
explained with the injected power of SN $\sim 10^{39}$ erg
s$^{-1}$.

In this letter we propose that the FB emission may result from star
capture processes, which have been developed by \citet{cheng1,cheng2} and
\citet{dog_pasj1,dog_pasj,dog_aa} to explain a wide range of X-ray and gamma-ray
emission phenomena from the GC.

\section{Observations}
The procedure of separation of the bubble emission from the total
diffuse emission of the Galaxy is described in \citet{meng}. It is important to note that the bubble
structure is seen when components of gamma-ray emission
produced by cosmic ray (CR) interaction with the background gas,
i.e. by CR protons ($\pi^o$ decay) and electrons
(bremsstrahlung) are removed.    \citet{meng} concluded the the
bubble emission was of the inverse Compton (IC) origin generated
by relativistic electrons.
Here we summarize the multi-wavelength observational constraints of the FB:
\begin{itemize}
\item The spectral shape and intensity of gamma-rays are almost
constant over the bubble region that suggests a uniform production
of gamma-rays in the FB. The total gamma-ray flux from the bubble at energies
$E_\gamma>1$ GeV is $F_\gamma\sim 4\times 10^{37}$ erg s$^{-1}$
and the photon spectrum of gamma-rays is power-law,
$dN_\gamma/dE_\gamma\propto E_\gamma^{-2}$ for the range 1-100
GeV (\citet{meng});
\item In the radio range the bubble is seen  from the tens GHz WMAP data
as a microwave residual spherical excess  \emph{("the
microwave haze")} above the GC   $\sim4$ kpc in radius
\citep{fink}. Its power spectrum in the frequency range 23 - 33
GHz is described as power-law, $\Phi_{\nu}\propto \nu^{-0.5}$. For
the magnetic field strength $H\sim
 10~\mu$G the energy range of electrons responsible for emitting these radio waves is within
the range 20 - 30 GeV and their spectrum is  $dN_e/dE_e\propto
E^{-2}_e$;
\item The ROSAT 1.5 keV  X-ray data clearly showed the
characteristic of a bipolar structure \citep{cohen} that
aligned well with the edges of the Fermi bubble. The ROSAT
structure is explained as due to a fast wind  which drove a shock
into the halo gas with the velocity $v_{sh}\sim 10^8$cm s$^{-1}$.
This phenomenon requires an energy release   of about $10^{55}$
ergs at the GC and  this activity should be periodic on a
timescale of $10-15$ Myr.
\item The similarities of the morphology of the radio, X-ray and
gamma-ray structures strongly suggest their common origin.
\end{itemize}

In the case of electron (leptonic) model of \citet{meng}
gamma-rays are produced by scattering of relativistic
electrons on background soft photons, i.e. relic, IR and optical
photons from the disk.

\section{The bubble origin in model of a multiple star capture by the central black hole}
In this section we present our ideas about the origin of the Fermi Bubble in the
framework of star capture by the central SMBH. The
process of gamma-ray emission from the bubble is determined by a
number of stages of energy transformation.  Each of these stages
actually involves complicated physical processes. The exact
details of these processes are still not understood very well.
Nevertheless, their qualitative features do not depend on these
details. In the following, we only briefly describe these
processes and give their qualitative interpretations. We begin to
describe processes of star capture by the central black hole as
presented in \citet{dog_aa}.

\subsection{Star capture by the central black hole}

As observations show, there is a supermassive black hole (Sgr
A$^\ast$) in the center of our Galaxy with a mass of $\sim
4\times10^6~M_{\odot}$ . A total tidal disruption of a star occurs
when the penetration parameter $b^{-1}\gg 1$, where $b$ is the
ratio of  the periapse distance $r_p$  to the tidal radius $R_T$.
The tidal disruption rate $\nu_s$ can be approximated to within an
order of magnitude $\nu_s\sim 10^{-4}-
10^{-5}$~yr$^{-1}$ \citep[see the review of][]{alex05}.

The energy budget of a tidal disruption event follows from
analysis of star matter dynamics.
Initially about 50\% of the stellar mass becomes tightly bound to
the black hole , while the remainder 50\% of the stellar mass is
forcefully ejected \citep[see, e.g.][]{ayal}. The kinetic energy
carried by the ejected debris is a function of the penetration
parameter $b^{-1}$ and can significantly exceed that released by a
normal supernova ($\sim 10^{51}$~erg) if the orbit is highly
penetrating \citep{alex05},
\begin{equation}\label{energy}
  W\sim 4\times 10^{52}\left(\frac{M_\ast}{M_\odot}\right)^2
  \left(\frac{R_\ast}{R_\odot}\right)^{-1}\left(\frac{M_{\rm bh}/M_\ast}{10^6}\right)^{1/3}
  \left(\frac{b}{0.1}\right)^{-2}~\mbox{erg}\,.
\end{equation}
Thus, the mean kinetic energy per escaping nucleon is estimated as
$E_{\rm esc}\sim 42 \left(\frac{\eta}{0.5}\right)^{-1} \left(\frac{M_\ast}{M_\odot}\right)
  \left(\frac{R_\ast}{R_\odot}\right)^{-1}\left(\frac{M_{\rm bh}/M_\ast}{10^6}\right)^{1/3}
  \left(\frac{b}{0.1}\right)^{-2}~\mbox{MeV}$,
where $\eta M_\ast$ is the mass of escaping material. From $W$ and
$\nu_s$ we obtain that the average power in the form of a flux of
subrelativistic protons. If $W\sim 3\times 10^{52}$ erg and $\nu_s\sim
3\times 10^{-5}$ yr$^{-1}$, we get $\dot{W}\sim 3\times 10^{40}$ erg s$^{-1}$.

In \citet{dog_pasj} we described the injection spectrum of protons
generated by processes of star capture as monoenergetic. This is a
simplification of the injection process because a stream of
charged particles  should be influenced by different plasma
instabilities,as it was shown by \citet{chech} for the case of
relativistic jets. At first  the jet material is moved by inertia.
Then due to the excitation of plasma instabilties in the flux, the
particle distribution functions, which were initially delta
functions both in angle and in energy, transforms into complex
angular and energy dependencies.

\subsection{Plasma heating by subrelativistic protons}

Subrelativistic protons lose
their energy mainly by Coulomb collisions, i.e.
$\frac{dE_p}{dt}=-\frac{4\pi ne^4}{m_e
\mathrm{v}_{\rm{p}}}\ln\Lambda$, where $\mathrm{v}_{\rm{p}}$ is
the proton velocity, and $\ln\Lambda$ is the Coulomb logarithm. In
this way the protons
 transfer almost all their energy to the background plasma and
heat it. This process was analysed in \citet{dog_pasj,dog_aa}.
  For the GC parameters the
average time of Coulomb losses for several tens MeV protons is
several million years. The radius of plasma heated by the protons
is estimated from the diffusion equation describing propagation
and energy losses of protons in the GC \citep[][]{dog_pasj2}. This
radius is  about 100 pc. The temperature of heated plasma is
determined by the energy which these protons transfer to the
background gas.  For $\dot{W}\sim 10^{40}-10^{41}$erg s$^{-1}$ the
plasma temperature is about 10 keV \citep{koyama07} just as
observed by Suzaku for the GC. The plasma with such a  high
temperature cannot be confined at the GC and therefore it expands
into the surrounding medium.

\subsection{The hydrodynamic expansion stage}
Hydrodynamics of gas expansion was described in many monographs
and reviews \citep[see e.g.][]{kogan}. As the time of star capture may be smaller than
the time of proton energy losses, we have almost stationary energy
release in the central region with a power $\dot{W}\sim 3\times
10^{40}$ erg s$^{-1}$. This situation is very similar to that
described by \citet{weav77} for a stellar wind expanding into a
uniform density medium. The model describes  a star at time t=0
begins to blow a spherically symmetric wind with a velocity of
stellar wind  $V_w$, mass-loss rate $dM_w/dt=\dot{M_w}$, and a
luminosity $L_w=\dot{M_w}V_w^2/2$ which is analogous to the power
$\dot{W}$ produced by star capture processes. Most of the time
of the evolution is occupied by the so-called snowplow phase when
a thin shock front is propagating through the medium . The
 shock is expanding as
\begin{equation}
R_{sh}(t)=\alpha \left(\frac{L_wt^3}{\rho_0}\right)^{1/5}
\label{rho}
\end{equation}
where $\rho_0=n_0m_p$ and $\alpha$ is a constant of order of
unity. The velocity distribution inside the expanding
region $u(z)$ is nonuniform.

Our extrapolation of this hydrodynamic solution onto the Fermi
bubble is, of course, rather rough. First, the gas distribution in
the halo is nonuniform. Second, the analysis does not take into
account particle acceleration by the shock. A significant fraction
of the shock energy is spent on acceleration that modify the shock
structure. Nevertheless, this model presents a qualitative picture
of a shock in the halo.

\subsection{Shock wave acceleration phase and non-thermal emission}

The theory of particle acceleration by shock is described in many
publications. This theory has been developed, and  bulky numerical
calculations have beeen performed to calculate  spectra of
particles accelerated by supernova (SN) shocks and emission
produced by accelerated electrons and protons in the range from
radio up to TeV gamma-rays \citep[see e.g.][]{volk}. Nevertheless
many aspects of these processes are still unclear. For instance,
the ratio of electrons to protons accelerated by shocks is not
reliably estimated \citep[see][]{byk}.

We notice that the energy of shock front expected in the GC is nearly two
orders of  magnitude larger than that of the energy
released in a SN explosion. Therefore, process of particle
acceleration in terms of sizes of envelope, number of accelerated
particles, maximum energy of accelerated particles, etc. may
differ significantly from those obtained for SNs. Below we present
simple estimations of electron acceleration by shocks.

The injection spectrum of electrons accelerated  in shocks is
 power-law, $Q(E_e)\propto
E_e^{-2}$, and the maximum energy of accelerated electrons can be
estimated from a kinetic equation describing their spectrum at the
shock \citep{ber90}, $E_{max}\simeq {v_{sh}^2}/{D\beta}$, where
$v_{sh}\sim 10^8$cm/s is the velocity of shock front, $D$ is the
diffusion coefficient at the shock front and the energy losses of
electrons (synchrotron and IC) are: $dE_e/dt=-\beta E_e^2$. Recall
$\beta$ is a function of the magnetic and background radiation
energy densities, $\beta \sim w\sigma_Tc/(m_ec^2)^2$, where
$\sigma_T$ is Thompson cross section and $w=w_{ph}+w_B$ is the
combined energy density of background photons $w_{ph}$ and the
magnetic energy density $w_B$ respectively. It is difficult to
estimate the diffusion coefficient near the shock. For qualitative
estimation, we can use the Bohm diffusion ($\sim r_L(E_e)$c),
where $r_L$ is the Larmor radius of electrons. Using the typical
values of these parameters, we obtain $E_{max}\sim
1~TeV~v_8B_{-5}^{1/2}w_{-12}^{-1/2}$, where $v_8$ is the shock
velocity in units of $10^8$cm/s, $B_{-5}$ is the magnetic field in
the shock in units of $10^{-5}$G and $w_{-12}$ is the energy
density in units of 10$^{-12}$erg/cm$^3$.

The spectrum of electrons in the bubble is modified
by processes of energy losses and escape. It can be derived from
the kinetic equation
\begin{equation}
\frac{d}{dE}\left(\frac{dE}{dt}N\right)+\frac{N}{T}=Q(E)
\end{equation}
where $dE/dt=\beta E^2+\nabla u(z)E$ describes the inverse
Compton, synchrotron and adiabatic (because of wind velocity
variations) energy losses, $T$ is the time of particle escape from
the bubble, and $Q(E)=KE^{-2}\theta(E_{max}-E)$ describes
particles injection spectrum  in the bubble. As one can see, in
general case the spectrum of electrons in the bubble cannot be
described by a single power-law as assumed by \citet{meng}. The
spectrum of electrons has a break at the energy $E_b\sim 1/\beta
T$ where  $T$ is the characteristic time
of either particle escape from the bubble or of the adiabatic
losses, e.g. for $\nabla u=\alpha$ the break position follows from
$T\sim 1/\alpha$. By solving equation 3, we can see that the electron spectrum cannot be described by a
single power-law even in case of power-law injection (see eg. Berezinskii et al., 1990).

\begin{figure}
\epsscale{.80} \plotone{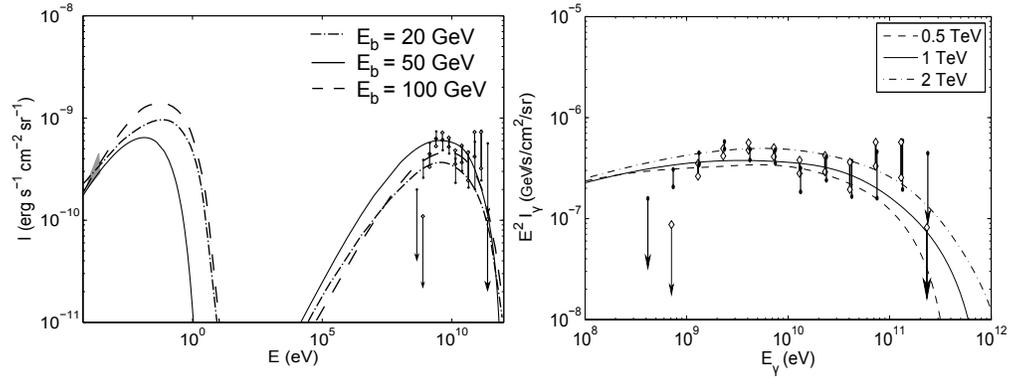} \caption{ (a)Energy fluxes produced by
Synchrotron and IC with $E_{max}$=1TeV and different $E_b$. (b)The gamma-ray
spectra with $E_b$=50GeV and different $E_{max}$.
\label{emission}.}
\end{figure}

The distribution of background photons can be derived  from
GALPROP program. The average energy density of background photons
in the halo are $w_o=2$ eV cm$^{-3}$ for optical and $w_{IR}=0.34$
eV cm$^{-3}$ for IR. These background soft photon energy densities
are obviously not negligible in comparing with $w_{CMB}=0.25$ eV
cm$^{-3}$ for the relic photons and also comparable with the
magnetic energy density ($\sim 1(H/5\times 10^{-6}G)^2
eV/cm^3$). The expected IC energy flux of gamma-rays and
synchrotron radiation emitted from the same population of
electrons described above are shown in Fig. \ref{emission} 
for different values of $E_b$ and $E_{max}$. The
Klein-Nishina IC cross-section \citep{gould} is used. The observed
spectrum of radioemission in the range 5-200GHz and gamma-rays are
taken from \citet{dobler1} and \citet{meng} respectively.
The inverse Compton gamma ray spectrum is formed by
scattering on three different components of the background photons.
When these three components are combined (see Fig.
\ref{emission}b), they mimic a photon spectrum $E_\gamma^{-2}$ and
describe well the data shown in Fig. 23 of \citet{meng}.
 We want to remark that although a single power law with the spectral
 indexes in between 1.8 and 2.4 in the energy range of electrons from 0.1 to 1000 GeV
 can also
explain both the Fermi data as well as the radio data as suggested
by Su et al. (2010). Theoretically a more complicated electron spectrum will be
developed when the cooling time scale is comparable with the
escape time even electrons are injected with a single power law as shown in equation 3.

\subsection{The thermal emission from heated plasma}

In our model there is 10keV hot plasma with power $\dot{W} \sim
3\times 10^{40}$erg/s injected into bubbles. Part of these
energies are used to accelerate the charged particles in the shock
but a good fraction of energy will be used to heat up the gas in
the halo due to Coulomb collisions. The temperature of halo gas
can be estimate as $nd^3kT\approx \dot{W}d/v_w$ which gives
$kT\sim 1.5 (\dot{W}/3\times
10^{40}erg/s)v_8^{-1}(d/5kpc)^{-2}(n/10^{-3})$keV. The thermal
radiation power from the heated halo gas is simply given by
$L_{th}=1.4\times 10^{-27}n_en_iZ^2T^{1/2}$erg/cm$^3$
\citep{rybicki}.
 By using kT=1.5keV, $n_e=n_i=10^{-3}cm^{-3}$ and Z=1, we find
$L_x\sim 10^{38}$erg/s.

\section{Discussion}
The observed giant structure of FB is difficult to be explained by other processes.
We suggest that periodic star capture processes by the central SMBH can inject
$\sim 3\times 10^{40}$erg/s hot plasma into the galactic halo. The hot gas can
expand hydrodynamically and form shock to accelerate electrons to relativistic
speed. Synchrotron radiation and inverse Compton scattering with the background
soft photons produce the observed radio and gamma-rays respectively. Acceleration of protons by the same shock
may contribute to Ultra-high energy cosmic rays, which will be considered in future works.

It is interesting to point out that the mean free path of TeV electrons
$\lambda \sim \sqrt{D/\beta E_e}\sim 50 D_{28}^{1/2}\tau_5^{1/2}$pc, where
$D_{28}$ and $\tau_5$ are the diffusion coefficient and cooling time for TeV
electrons in units of $10^{28}cm^2/s$ and $10^5$yrs respectively. This estimated
mean free path is much shorter than
the size of the bubble. In our model the capture time is once every $\sim 3\times 10^4$yr,
we expect that there is about nearly 100 captures in 3 million years. Each of these
captures can produce an individual shock front, therefore the gamma-ray radiation can
be emitted uniformly over the entire bubble.

Furthermore we can estimate the shape of the bubble, if we simplify
the geometry of our model as follows. After each capture a disk-like
hot gas will be ejected from the GC. Since the gas pressure in the halo
($n(r)kT\sim 10^{-14}(n/3\times10^{-3}cm^{-3})(T/3\times 10^4K)$erg/cm$^3$) is low and decreases quadratically
for distance larger than 6kpc(Paczynski 1990),
we can assume that the hot gas can escape freely vertically, which is defined as the z-direction and hence the z-component of the wind velocity
$v_{wz} = constant$ or $z=v_{wz}t$. The ejected disk has a thickness $\Delta z=v_{wz} t_{cap}$,
where $t_{cap}=3\times 10^4$yr is the
capture time scale. On the other hand the hot gas can also expand laterally and its radius along the direction of the galactic disk is given by
$x(t)=v_{wx} t + x_0\approx v_{wx} t$, where $x_0\sim 100pc$(cf. section 3.2). When the expansion speed is supersonic then
shock front can be formed at the edge of the ejected disk. In the vertical co-moving frame of the ejected disk
the energy of the disk is
$\Delta E$, which is approximately constant if the radiation loss is small. The energy
conservation gives $\Delta E={1\over2}m v_{wx}^2$ with $m=m_0+m_s(t)=m_0+\pi x^2\Delta
z\rho$, where $m_0\approx 2\Delta E/v_w^2$ is the initial mass in the ejected disk, $m_s$ is the swept-up mass from the surrounding gas when the disk is expanding laterally, and  $\rho=m_pn$ is the density of the medium
surrounding the bubble. Combing the above equations, we can obtain
$\Delta E={1\over2}[m_0+\pi (v_{wx}t)^2\Delta z\rho]
v_{wx}^2$.
There are two characteristic stages, i.e. free expansion stage, in which $v_{wx}\approx $constant for $m_0>m_s(t)$ and deceleration stage for $m_0<m_s(t)$. The time scale switching from free expansion to deceleration is given by $m_0=m_s(t_s)$ or $t_s=\sqrt{m_0/\pi \Delta z \rho v_{wx}^2}$. In the free expansion stage, we obtain
$x=\frac{v_{wx}}{v_{wz}}z \sim z$
for $x<v_{wx} t_s=x_s$.
In the deceleration stage, $\Delta E\approx{1\over2}\pi (v_{wx}t)^2\Delta z\rho
v_{wx}^2$, we obtain
$(x/v_w t_s)=\left(\frac{\Delta E}{\pi t_s^2 v_{wx}^4 \Delta z \rho}\right)^{1/4}$$\left(\frac {z}{v_w t_s}\right)^{1/2}$$
\approx 0.9 \left(\frac {z}{v_w t_s}\right)^{1/2}$, we have approximated $v_{wx}\sim v_{wz}\sim v_w\sim 10^8$cm/s, $\Delta E=3\times 10^{52}$ergs and $\rho/m_p=3\times 10^{-3}cm^{-3}$.
The switching from a linear relation to the quadratic relation takes place at $z_s\sim v_w t_s \sim 300$pc. The quasi-periodic injection of disks into the halo can form a sharp edge, where shock fronts result from the laterally expanding disks with quadratic shape, i.e. $z\sim x^2$. In fitting the gamma-ray spectrum it gives $E_b \sim 50$GeV, which corresponds to a characteristic time scale of either adiabatic loss or particle escape $\sim 15$Myrs. By using equation 2, the characteristic radius of FB is about 5kpc, which is quite close to the observed size of FB.

\acknowledgments We thank the anonymous referee for very useful suggestions and Y.W. Yu for useful discussion.
KSC is supported by a grant under HKU
7011/10p. VAD and DOC are partly supported by
the NSC-RFBR Joint Research Project RP09N04
and 09-02-92000-HHC-a.  CMK is supported in part by the National Science
Council, Taiwan, under grants NSC 98-2923-M-008-001-MY3 and NSC
99-2112-M-008-015-MY3.





\begin{thebibliography}{}
\bibitem[Aharonian et al.(2006)]{ahar}
Aharonian, F., Akhperjanian, A. G., Bazer-Bachi, A. R.  et al.
2006, Nature, 439, 695
\bibitem[Alexander(2005)]{alex05}
Alexander, T. 2005, PhR, 419, 65
\bibitem[Ayal et al.(2000)]{ayal}
Ayal, S., Livio, M., and  Piran, T. 2000, \apj, 545, 772
\bibitem[Berezhko \& Voelk(2010)]{volk}
Berezhko, E. G.; Voelk, H. J. 2010, A\&A, 511, 34
\bibitem[Berezinskii et al.(1990)]{ber90}
Berezinskii, V. S., Bulanov, S. V., Dogiel, V. A., Ginzburg, V.
L., and Ptuskin, V. S. 1990, {\it Astrophysics of Cosmic Rays},
ed. V.L.Ginzburg, (Norht-Holland,
Amsterdam)
\bibitem[Bisnovatyi-Kogan \& Silich(1995)]{kogan}
Bisnovatyi-Kogan, G. S., \& Silich, S. A.  1995, RvMP, 67, 661
\bibitem[Blumenthal \& Gould(1970)]{gould}
Blumenthal, G. R., \& Gould, R. J. 1970, RvMP, 42, 237
\bibitem[Bogdan \& Gilfanov(2010)]{gilf}
Bogdan, A., \& Gilfanov, M. 2010, MNRAS, 405, 209

\bibitem[Bland-Hawthorn \& Cohen(2003)]{cohen}
Bland-Hawthorn, J. \& Cohen, M. 2003, ApJ, 582, 246

\bibitem[Bykov \& Uvarov(1999)]{byk}
Bykov, A. M., \& Uvarov, Yu. A. 1999, JETP, 88, 465
\bibitem[Cheng et al.(2006)]{cheng1}
Cheng, K.-S., Chernyshov, D. O., and Dogiel, V. A. 2006, \apj,
645, 1138.
\bibitem[Cheng et al.(2007)]{cheng2}
Cheng, K.-S., Chernyshov, D. O., and Dogiel, V. A. 2007, \aap,
473, 351.
\bibitem[Crocker \& Aharonian(2010)]{crocker}
Crocker, R., \& Aharonian, F. 2010, arXiv1008.2658
\bibitem[Crocker et al.(2010a)]{crocker2}
Crocker, R. M., Jones, D. I., Aharonian, F. et al. 2010a, MNRAS,
411, L11
\bibitem[Crocker et al.(2010b)]{crocker1}
Crocker, R. M., Jones, D. I., Aharonian, F. et al. 2010b,
arXiv1011.0206
\bibitem[Diehl et al.(2006)]{diehl}
Diehl, R., Prantzos, N., \& von Ballmoos, P. 2006, Nuclear Physics
A, 777, 70
\bibitem[Dobler \& Finkbeiner(2008)]{dobler1}
Dobler, G., Finkbeiner, D. P. 2008, ApJ, 680, 1222

\bibitem[Dobler et al.(2010)]{dobler2}
Dobler, G. \& Finkbeiner, D. P., Cholis, I. et al. 2010, ApJ, 717,
825
\bibitem[Dogiel et al.(2009a)]{dog_pasj1}
Dogiel, V.,  Cheng, K.-S.,  Chernyshov D. et al.   2009a,  PASJ,
61, 901
\bibitem[Dogiel et al.(2009b)]{dog_pasj2}
Dogiel, V.,  Chernyshov D., Yuasa, T. et al.   2009b, 61, PASJ,
1093

\bibitem[Dogiel et al.(2009c)]{dog_pasj}
Dogiel, V.,  Chernyshov D., Yuasa, T. et al.   2009c, 61, PASJ,
1099
\bibitem[Dogiel et al.(2009d)]{dog_aa}
Dogiel, V. A., Tatischeff, V., Cheng, K.-S. et al. 2009d, A\&A,
508, 1
\bibitem[Finkbeiner(2004)]{fink}
Finkbeiner, D. P. 2004, ApJ, 614, 186
\bibitem[Ginzburg et al.(2004)]{chech} Ginzburg, S. L.,
D'Yachenko, V. F., Paleychik, V. V., Sudarikov, A. L., \&
Chechetkin, V. M. 2004, Astronomy Letters, 30, 376
\bibitem[Inui et al.(2009)]{inui}
Inui, T., Koyama, K., Matsumoto, H., \& Tsuru, T.G. 2009, PASJ,
61, S241
\bibitem[Knoedlseder et al.(2005)]{knoed}
Knoedlseder, J., Jean, P., Lonjou, V. et al. 2005, A\&A, 441, 513
\bibitem[Koyama et al.(2007)]{koyama07}
Koyama, K., Hyodo, Y., Inui, T. et al. 2007, PASJ, 59, 245

\bibitem[Paczynski(1990)]{paczynski}
Paczynski, B. 1990, ApJ, 348, 485
\bibitem[Ponti et al.(2010)]{ponti}
Ponti, G., Terrier, R., Goldwurm, A. et al. 2010,  ApJ,  714, 732
\bibitem[Rybicki \& Lightman(1979)]{rybicki}
Rybicki, G.B. \& Lightman, A.L. 1979, "Radiative Processes in
Astrophysics", New York: Wiley
\bibitem[Su et al.(2010)]{meng}
Su, M., Slatyer, T. R., \& Finkbeiner, D. P. 2010, ApJ, 724, 1044
\bibitem[Terrier et al.(2010)]{terr}
Terrier, R., Ponti, G., Belanger, G. et al. 2010, ApJ, 719, 143
\bibitem[Weaver et al. (1977)]{weav77}
Weaver, R., McCray, R., Castor, J., Shapiro, P., and Moore, R.
1977, ApJ, 218, 377
\end{thebibliography}
\end{document}